\documentclass[a4paper,fleqn,usenatbib]{mnras}

\usepackage[T1]{fontenc}
\usepackage{ae,aecompl}

\usepackage{graphicx}	
\usepackage{amsmath}	
\usepackage{amssymb}	
\usepackage{hyperref}    

\title[Two populations in NGC~121]{The Search for Multiple Populations in Magellanic Cloud Clusters I: Two Stellar Populations in the Small Magellanic Cloud Globular Cluster NGC~121\thanks{Based on observations made with the NASA/ESA Hubble Space Telescope, and obtained from the Hubble Legacy Archive, which is a collaboration between the Space Telescope Science Institute (STScI/NASA), the Space Telescope European Coordinating Facility (ST-ECF/ESA) and the Canadian Astronomy Data Centre (CADC/NRC/CSA).}}

\author[F. Niederhofer et al.]{
F. Niederhofer$^{1}$\thanks{FN: fniederhofer@stsci.edu}, 
N. Bastian$^{2}$,
V. Kozhurina-Platais$^{1}$,
S. Larsen$^{3}$,
M. Salaris$^{2}$,
\newauthor
E. Dalessandro$^{4}$,
A. Mucciarelli$^{4}$,
I. Cabrera-Ziri$^{2,5}$,
M. Cordero$^{6}$,
D. Geisler$^{7}$,
M. Hilker$^{5}$,
\newauthor
K. Hollyhead$^{2}$,
N. Kacharov$^{8}$,
C. Lardo$^{2}$,
C. Li$^{9}$,
D. Mackey$^{10}$,
and I. Platais$^{11}$
\\
$^{1}$ Space Telescope Science Institute, 3700 San Martin Drive, Baltimore, MD 21218, USA \\
$^{2}$ Astrophysics Research Institute, Liverpool John Moores University, 146 Brownlow Hill, Liverpool L3 5RF, UK \\
$^{3}$ Department of Astrophysics/IMAPP, Radboud University, P.O. Box 9010, 6500 GL Nijmegen, The Netherlands \\
$^{4}$ Department of Physics and Astronomy, University of Bologna, Viale Berti Pichat 6/2, I-40127 Bologna, Italy \\
$^{5}$ European Southern Observatory, Karl-Schwarzschild-Stra\ss e 2, D-85748 Garching bei M\"unchen, Germany \\
$^{6}$ Astronomisches Rechen-Institut, Zentrum f\"ur Astronomie der Universit\"at Heidelberg, M\"onchhofstra\ss e 12-14, D-69120 Heidelberg, Germany \\
$^{7}$ Departamento de Astronomia, Universidad de Concepcion, Casilla 160-C, Chile \\
$^{8}$ Max-Planck-Institut f\"ur Astronomie, K\"onigstuhl 17, D-69117 Heidelberg, Germany \\
$^{9}$ Department of Physics and Astronomy, Macquarie University, Sydney, NSW 2109, Australia \\
$^{10}$ Research School of Astronomy and Astrophysics, Australian National University, Canberra, ACT 2611, Australia \\
$^{11}$ Department of Physics and Astronomy, Johns Hopkins University, 3400 North Charles Street, Baltimore, MD 21218, USA
}

\date{Accepted XXX. Received YYY; in original form ZZZ}

\pubyear{2016}

\begin{document}
\label{firstpage}
\pagerange{\pageref{firstpage}--\pageref{lastpage}}
\maketitle

\begin{abstract}
We started a photometric survey using the WFC3/UVIS instrument onboard the Hubble Space Telescope to search for multiple populations within Magellanic Cloud star clusters at various ages. In this paper, we introduce this survey. As first results of this programme, we also present multi-band photometric observations of NGC~121 in different filters taken with the WFC3/UVIS and ACS/WFC instruments. We analyze the colour-magnitude diagram (CMD) of NGC~121, which is the only "classical" globular cluster within the Small Magellanic Cloud. 
Thereby, we use the pseudo-colour C$_{F336W,F438W,F343N}=(F336W-F438W)-(F438W-F343N)$ to separate populations with different C and N abundances. We show that the red giant branch splits up in two distinct populations when using this colour combination. NGC~121 thus appears to be similar to Galactic globular clusters in hosting multiple populations. The fraction of enriched stars (N rich, C poor) in NGC~121 is about 32\%$\pm$3\%, which is lower than the median fraction found in Milky Way globular clusters. The enriched population seems to be more centrally concentrated compared to the primordial one. These results are consistent with the recent results by Dalessandro et al. (2016). The morphology of the Horizontal Branch in a CMD using the optical filters $F555W$ and $F814W$ is best produced by a population with a spread in Helium of $\Delta Y$=0.025$\pm$0.005.
\end{abstract}

\begin{keywords}
galaxies: star clusters: individual: NGC~121 -- galaxies: individual: SMC -- Hertzsprung--Russell and colour--magnitude diagrams
\end{keywords}



\section{Introduction}
\label{sec:intro}

A nearly ubiquitous property of ancient globular clusters (GCs) so far studied is that they host multiple populations in the form of internal chemical abundance variations in light elements \citep[see e.g.][for a review]{Gratton12}. So far, Ruprecht 106 and IC~4499 seem to be the only exceptions (see \citealt{Villanova13} and \citealt{Walker11}). Interestingly, these variations, which are not observed among field stars of the same metallicity, show correlated patterns in certain elements, e.g. the prominent Na-O anti-correlation \citep[e.g.][]{Carretta09} or the C-N anti-correlation \citep[e.g.][]{Cannon98}. These chemical anomalies are not only detected in Milky Way GCs but also in old clusters in nearby dwarf galaxies, like in the Fornax dwarf spheroidal galaxy \citep{Larsen14}, the Sagittarius dwarf galaxy \citep{Carretta10a, Carretta14} or the Large Magellanic Cloud \citep[LMC,][]{Mucciarelli09}.

The Na-O and C-N anti-correlations are ideal tracers of the multiple populations in GCs. These can be detected by spectroscopic analysis of individual stars in a cluster \citep[e.g.][]{Carretta09, Carretta15, Marino16} down to the main sequence \citep[e.g.][]{Harbeck03, D'Orazi10}. \citet{Marino08} in their study of the GC M4, combined spectroscopic and photometric data and showed that there is a direct relation between the broadening of the red giant branch (RGB) in the colour-magnitude diagram (CMD) and the spectroscopically determined populations with varying Na and O abundances (see \citealt{Dalessandro14} and \citealt{Mucciarelli16} for a similar study on NGC~6362). The different element abundances within the stars influence the colour of the stars in certain filter bands and can therefore result in a broadening or splitting of the various stellar evolutionary stages in the CMD, like the RGB, sub-giant branch (SGB) and the main sequence (MS). With the photometric precision of the Hubble Space Telescope (HST) combined with the usage of ultraviolet filters it is possible to trace the multiple populations throughout the entire CMD down to the lowest magnitudes \citep[e.g.][]{Milone15a, Piotto15}. As multiple populations are even observed along main sequence stars indicates that the formation mechanism must have acted already at early stages of the cluster's life.

As the observed multiple populations clearly contradict our view of star clusters as simple stellar populations, several scenarios have been put forward to explain this phenomenon in recent years. Most of them involve the formation of more than one generation of stars where the younger stars form out of a mix of pristine gas and the enriched ejected material from stars of the older generations. The different scenarios propose various types of polluter stars: interacting massive binaries \citep{deMink09}, fast rotating massive stars \citep[e.g.][]{Decressin09, Krause13} and asymptotic giant branch (AGB) stars \citep[e.g.][]{D'Ercole08}. Alternatively, \citet{Bastian13a} proposed the early disk accretion scenario where the accretion disks of low-mass pre-MS stars sweep up enriched material ejected by rotating massive stars of the same generation. However, all proposed scenarios have severe difficulties accounting for the breadth of observations \citep[see e.g.][]{Bastian15a, Renzini15}.
In order to produce the observed anti-correlations in certain elements and the fraction of enriched stars, the GCs must have been at least one order of magnitude more massive at birth in scenarios invoking multiple star formation epochs \citep[e.g.][]{D'Ercole08, Bekki11}. This is referred to as the "mass-budget problem" (see e.g. \citealt{Larsen12}, \citealt{BastianLardo15} and \citealt{Cabrera-Ziri15} for a discussion). Additionally, none of the proposed sources of enriched material is able to reproduce consistently the extent of the observed abundance patterns in GCs \citep{Bastian15b}.  

As the proposed theories do not require any specific conditions for the formation of multiple populations, they should also be present in younger clusters with comparable observed properties to ancient GCs. Several studies aimed to find indications of multiple populations or multiple star formation events in such clusters. However, up to now no clear evidence is found for either of these indicators in young clusters. \citet{Bastian13b} did not detect any signs of ongoing star formation in a sample of 130 massive ($10^4$ - $10^8$~M$_{\sun}$) clusters with ages between 10~Myr and 1~Gyr. Also numerous searches for age spreads in  extragalactic young massive clusters remain without any detection \citep[e.g.][]{Larsen11, BastianSilvaVilla13, Cabrera-Ziri14, Niederhofer15, Cabrera-Ziri16a}. In order to form a second generation of stars, a cluster either has to retain gas that is left over from the first star formation event or (re-)accrete fresh gas from its surroundings. But young clusters seem to remove the gas very efficiently already at young ages \citep[e.g.][]{Bastian14, Hollyhead15} and no clusters have been detected with an associated gas reservoir sufficient for a subsequent period of star formation \citep[e.g.][]{Cabrera-Ziri15, BastianStrader14, Longmore15}

In a series of papers, \citet{Mucciarelli08, Mucciarelli11, Mucciarelli14} spectroscopically studied RGB stars in the intermediate-age (1-3~Gyr) LMC clusters NGC 1651, NGC~1783, NGC~2173, NGC~1978 and NGC~1806, as well as in the $\sim$200~Myr old cluster NGC~1866, in large part motivated by the search for multiple populations. Among their sample of stars they did not detect any significant spread in light element abundances. \citet{Davies09} analyzed two Scutum Red Supergiant Clusters in the Milky Way, RSGC1 and RSGC2 ($\sim$2$\times$10$^4$~M$_{\sun}$) and found that they are chemically homogeneous. Similarly, \citet{Cabrera-Ziri16b} found that the young ($\sim$15~Myr) massive ($\sim$10$^6$~M$_{\sun}$) cluster NGC~1705:~1 shows [Al/Fe] abundances comparable to those of Small Magellanic Cloud (SMC) field red supergiant stars at the same metallicity, while an Al enhancement is generally observed in GCs showing multiple populations, although the authors could not rule out that small Al spreads were present.

The above-mentioned results challenge the interpretation that young massive clusters form the same way as GCs and may suggest that correlated anomalies in light elements are exclusively found in old GCs. However, due to the still small number of studies the hypothesis that young massive clusters are real counterparts of ancient GCs can not conclusively be discarded. 

We recently started a photometric survey of star clusters spanning a wide range of masses and ages within the Magellanic Clouds using the HST WFC3/UVIS instrument. This survey will help to answer the open question as to whether the age or the mass of a cluster is the critical parameter that determines if a cluster can host multiple populations. We included in our sample also NGC~121, the only "classical" GC in the SMC with an age $>$10~Gyr \citep{Glatt08a} as a benchmark object to test our methods. 
Note that, although NGC~121 is indeed the oldest SMC cluster, its age of about 10.5~Gyr is substantially younger than that of typical Milky Way or LMC globulars. In this paper, we report on the ability of the combination of the $F336W$, $F343N$, and $F438W$ filters to separate populations with different chemical abundances in the CMD. Furthermore, we apply our method to NGC~121 which is shown to host multiple populations as well \citep{Dalessandro16}.

The paper is structured as follows: In $\S$ \ref{sec:survey} we introduce our survey of Magellanic Cloud clusters. We describe the observations of NGC~121 and the data reduction procedure in $\S$ \ref{sec:obs}. The analysis of the data and the results are shown in $\S$ \ref{sec:analysis}. In $\S$ \ref{sec:conclusions} we discuss our results and draw final conclusion.

\section{The Survey}
\label{sec:survey}

\subsection{The Observations}

We started a photometric survey (GO-14069, PI. N. Bastian) using the WFC3/UVIS instrument onboard HST to search for multiple populations within massive star clusters of various ages. The main goal of this survey is to answer the question whether chemical variations within clusters are exclusively found in ancient (ages $\gtrsim$10~Gyr) GCs. We are imaging in total a sample of 12 clusters both in the LMC and SMC. We choose clusters with masses $\gtrsim 10^5$~M$_{\sun}$, so their masses are comparable to the masses of Galactic GCs that show multiple populations. Additionally, the clusters in our sample span a wide range of ages, going from $\sim$100~Myr to $>$ 10~Gyr. In Table~\ref{tab:cluster_list} we list the name of the clusters, their literature ages and masses, the filters in which the clusters have already been observed and the filters included in our survey.

Within this program, we are exploring the clusters in the ultraviolet/blue filters $F336W$, $F343N$, and $F438W$.  These specific filters have strong absorption lines of NH, CN and CH within their pass-bands which allows us to trace multiple populations in the CMDs of the target clusters. Several clusters in our sample have already available data in the $F336W$ and $F438W$/$F435W$ filters and we will add new observations in the $F343N$ filter. The observations in the ultraviolet/blue spectral range will be combined with already existing data from ACS/WFC and WFC3/UVIS in optical and infrared filters.  

\begin{table*} 
\centering
\caption{List of clusters in the survey\label{tab:cluster_list}}
\begin{tabular}{l c c c c c c c} 

\hline\hline
\noalign{\smallskip}
Cluster Name & Galaxy & Age & Ref. & Mass & Ref. & Existing Data & Filters added
\\
& & [Gyr] & & [10$^5$~M$_{\sun}$] & & & in this programme
\\
\noalign{\smallskip}
\hline
\noalign{\smallskip}
NGC~1850 & LMC & 0.1 & (1) & $\sim$2.0 & (1) & $-$ &  $F343N$, $F336W$, $F438W$
\\
NGC~1866 & LMC & 0.18 & (2) & $\sim$1.0 & (2) & $-$ & $F343N$, $F336W$, $F438W$
\\
NGC~1856 & LMC & 0.28 & (3) & $\sim$1.0 & (2) & $F336W$, $F438W$, $F814W$ & $F343N$
\\
NGC~419 & SMC & 1.2-1.6 & (4) & 2.4 & (8) & $F336W$, $F555W$, $F814W$ & $F343N$, $F438W$
\\
NGC~1783 & LMC & 1.75 & (5) & 2.6 & (8) & $F336W$, $F435W$, $F814W$ & $F343N$
\\
NGC~1806 & LMC & 1.70 & (5) & 1.3 & (8) & $F336W$, $F435W$, $F814W$ & $F343N$
\\
NGC~1846 & LMC & 1.75 & (5) & 1.7 & (8) & $F336W$, $F435W$, $F814W$ & $F343N$
\\
NGC~416 & SMC & 6.0 & (4) & 1.6 & (9) & $F555W$, $F814W$ & $F343N$, $F336W$, $F438W$
\\
NGC~339 & SMC & 6.0 & (4) & 0.8 & (9) & $F555W$, $F814W$ & $F343N$, $F336W$, $F438W$
\\
Lindsay~1 & SMC & 7.5 & (4) & $\sim$2.0 & (10) & $F555W$, $F814W$ & $F343N$, $F336W$, $F438W$
\\
NGC~361 & SMC & 7.9$^{a}$ & (6) & 2.0 & (9) & $-$ & $F343N$, $F336W$, $F438W$
\\
NGC~121 & SMC & 10.5 & (7) & 3.7 & (9) & $F336W$, $F438W$, $F555W$ & $F343N$
\\
\noalign{\smallskip}
\hline
\end{tabular}
\\
(1)~\citet{Niederhofer15}; 
(2)~\citet{BastianSilvaVilla13}; 
(3)~\citet{Milone15b}; 
(4)~\citet{Glatt08b}; 
(5)~\citet{Niederhofer16}; 
(6)~\citet{Mighell98}; 
(7)~\citet{Glatt08a}; 
(8)~\citet{Goudfrooij14}; 
(9)~\citet{McLaughlin05}; 
(10)~\citet{Glatt11}
\\
$^{a}$The age could be as low as 6 Gyr
\end{table*}

\begin{figure}
  \includegraphics[width=\columnwidth]{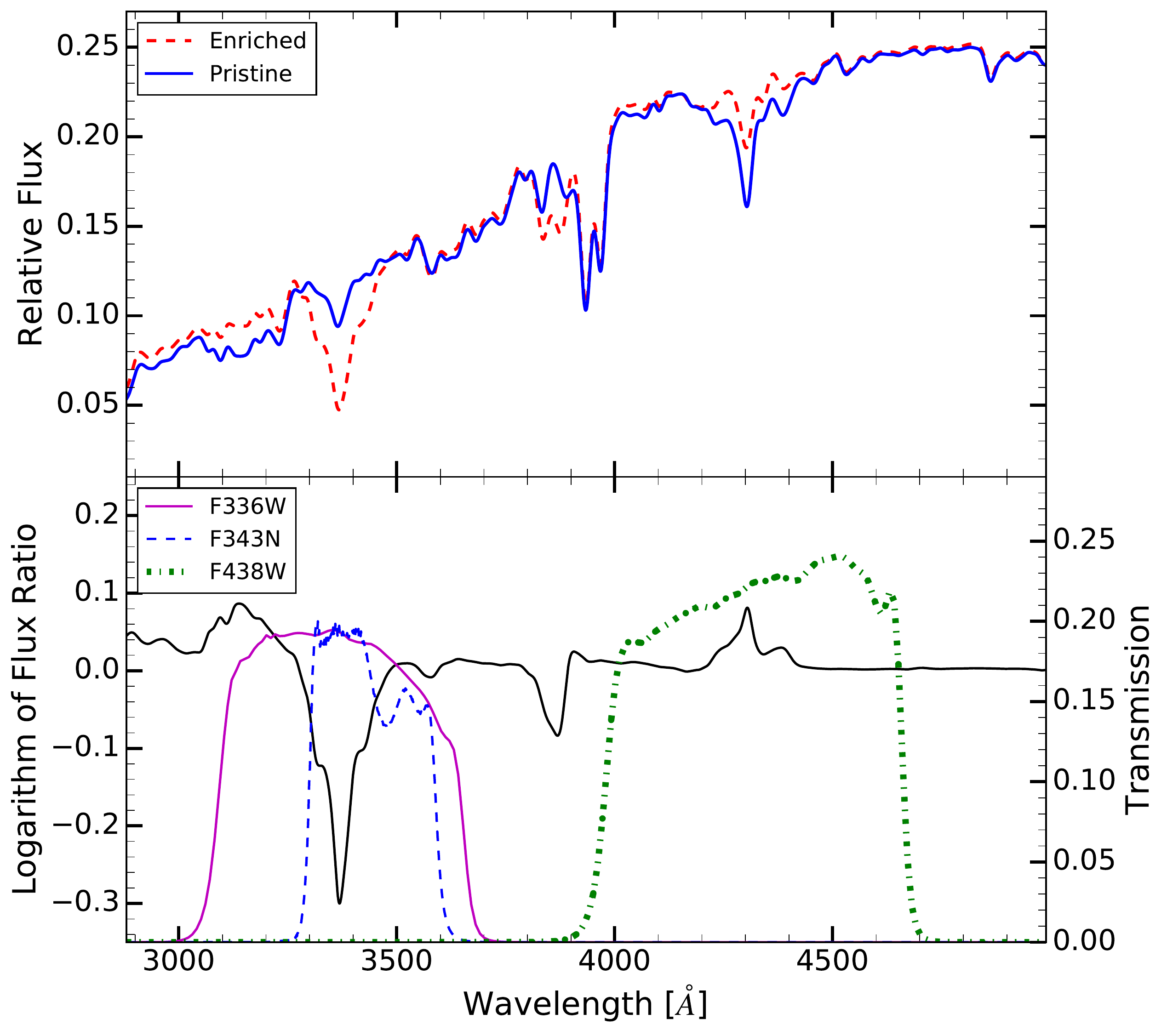} 
  \caption{\textbf{Upper Panel:} Model spectra of a typical RGB star in a 10~Gyr old population, with an effective temperature T$_{\rm{eff}}$ = 5220~K, surface gravity log~(g) = 2.71~dex, and metallicity [Fe/H] = $-$1.5~dex. The blue solid curve belongs to a star with a pristine composition, whereas the red dashed line corresponds to a chemically enriched star. \textbf{Lower Panel:} Logarithmic ratio of the fluxes of the enriched and the pristine star (black solid line) together with the transmission curves of the $F336W$ (purple solid line), $F343N$ (blue dashed line) and $F438W$ (green dash-dotted line) filters. The comparison is shown in the restframe wavelength, however the wavelength shift due to the relative motion of NGC~121 is $\sim$2\AA\ and can therefore be neglected.}
   \label{fig:spectra}
\end{figure}

\subsection{The Choice of the Filters}

The power of using a combination of ultraviolet filters to separate multiple populations in CMDs has already been demonstrated by the studies of, for example, \citet{Milone12} and \citet{Piotto15}. In this work, we use two wideband filters $F336W$ and $F438W$ together with the narrowband filter $F343N$. The upper panel of Figure \ref{fig:spectra} shows model spectra of a typical RGB star in a 10 Gyr old population with a metallicity [Fe/H] of $-1.50$~dex. The blue spectrum corresponds to a first population star, i.e. with primordial abundance pattern, whereas the red dashed line shows the spectrum of a second population star, i.e. enriched in N and Na and depleted in C and O. For the model spectra, we used the ATLAS12 and SYNTHE model atmosphere and spectral synthesis codes \citep{Sbordone04, Kurucz05}. We assumed an alpha-enhanced composition of $[\alpha/\mathrm{Fe}]=+0.4$ for the primordial population, and for the enriched population we used the "CNONa1" mixture of \citet{Sbordone11}. Compared to the primordial composition, the enriched model is enhanced in N by 1.8~dex and in Na by 0.8~dex and depleted in O by 0.8~dex and in C by 0.6~dex. The level of enrichment in this model is typical for Galactic GCs. We also computed a model with an intermediate composition where N and Na are enriched by 0.9~dex and 0.4~dex, whereas O and C are depleted by 0.4~dex and 0.6~dex, respectively. Note that these models do not take into account any (unknown) enhancement of the He abundance in the enriched stars. In the lower panel of Figure \ref{fig:spectra} we show the flux ratio of the second and first population star as a black line together with the transmission curves\footnote{\url{http://www.stsci.edu/hst/wfc3/ins_performance/throughputs/Throughput_Tables}} of the $F336W$, $F343N$, and $F438W$ filters (cf Figure 32 in \citealt{Milone12} and Figure 1 in \citealt{Piotto15} for similar plots). The $F336W$ and $F343N$ filters contain a strong NH absorption band at $\sim$3370\AA\ within their bandpasses, resulting in a drop of the flux ratio at these wavelengths. The $F438W$ filter is centred at the CH feature at $\sim$4300\AA\ which increases the flux ratio in this filter. Therefore, a combination of these three filters will separate first population stars (primordial) from second population stars (N rich, C poor) in the CMD. 

\begin{figure}
\centering
\includegraphics[width=\columnwidth]{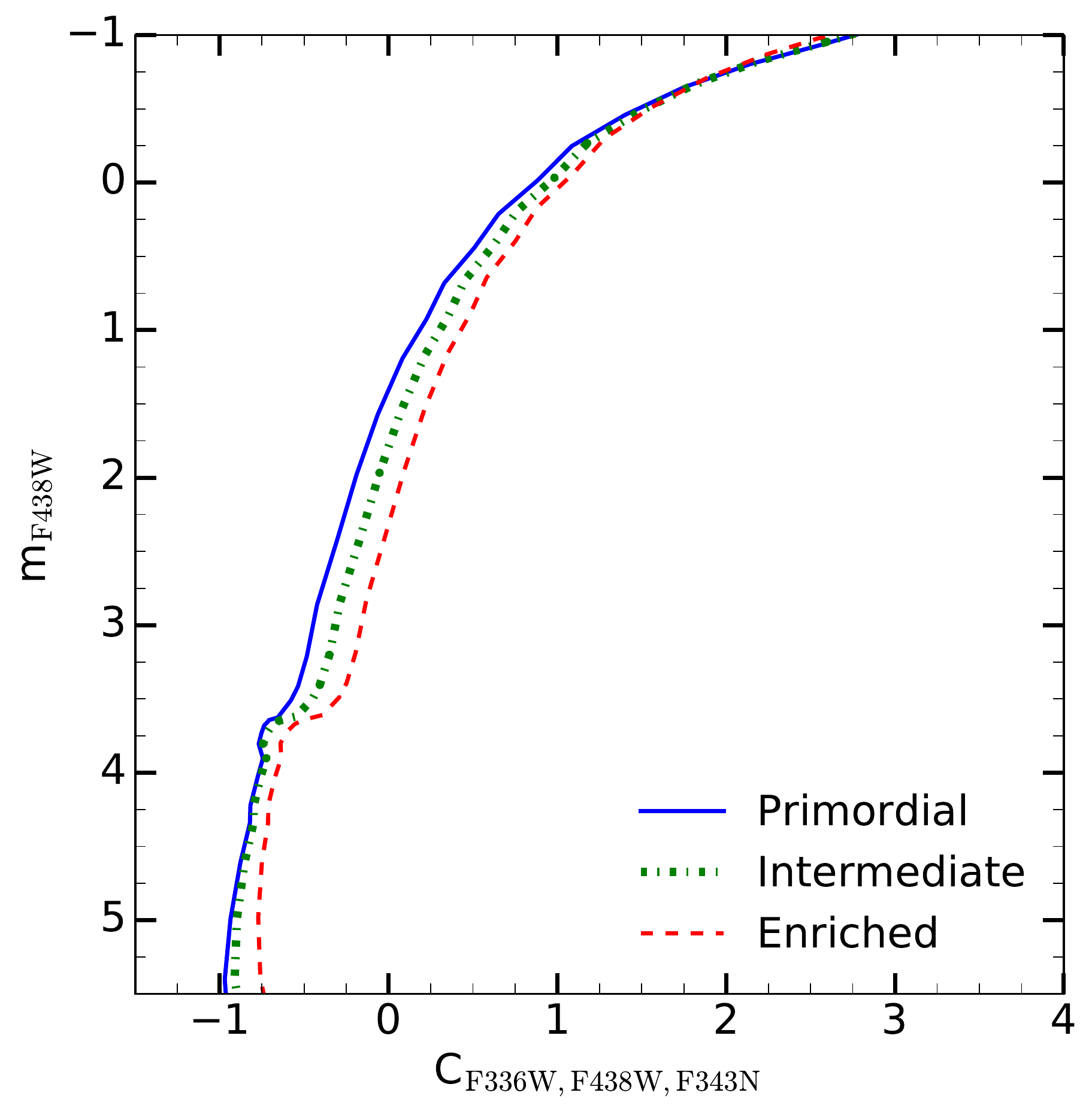}
\caption{Isochrones of a 10 Gyr old stellar population in the $m_{F438W}$ vs. $C_{F336W,F438W,F343N}$ CMD, showing the upper part of the main sequence as well as the RGB. The blue solid line corresponds to a population with primordial composition, whereas the green dash-dotted line shows stars with an intermediate enriched composition and the red dashed line shows enriched stars (see text for more details).}
\label{fig:isoc}
\end{figure}

We found a pseudo colour of the form $(F336W-F438W)-(F438W-F343N)$ (hereafter $C_{F336W,F438W,F343N}$) as the CMD's x-axis as an ideal combination for uncovering multiple populations with these filters. This is demonstrated in Figure \ref{fig:isoc} which shows the splitting of theoretical isochrones of a primordial (blue solid), intermediate (green dash-dotted) and enriched (red dashed) 10 Gyr old population. The separation of the isochrones is most evident in the lower part of the RGB where the primordial and enriched isochrones are about 0.2 to 0.3 mag in  $C_{F336W,F438W,F343N}$ apart from each other which is easily detectable. To compute the model colours for the various compositions, we used the spectra described above together with the most recent isochrones from the Padova website \citep{Bressan12, Chen14, Chen15, Tang14}.

\section{Observations and Data Reduction}
\label{sec:obs}

\subsection{Observations}
\label{subsec:obs}

The data of NGC~121 used in this paper comprise three sets of observations. 
The first set of observations are data from the ACS/WFC instrument in the $F555W$ and $F814W$ filters (GO-10396, PI. J.Gallagher) that have been taken in January 2006. The second set consists of archival observational data taken in 2014 with the WFC3/UVIS camera onboard HST through the $F336W$, $F438W$ and $F814W$ filters (GO-13435, PI. M. Monelli). The third set consists of a part of the current program GO-14069 (PI. N. Bastian) where NGC~121 is observed with the WFC3/UVIS instrument using the $F343N$ narrow-band filter. The exposure times were varied between short, intermediate and long exposures (see Table \ref{tab:obs_log} for a detailed journal of the different observations).

\begin{table*} 
\centering
\caption{Journal of the NGC~121 Observations\label{tab:obs_log}}
\begin{tabular}{l c c c c c c c} 

\hline\hline
\noalign{\smallskip}
Proposal ID & Date & Filter & Instrument & Exposure time & RA & Dec & HST Roll-angle PA\_V3
\\
& (yyyy-mm-dd) & & & (s) & & & (degree)
\\
\noalign{\smallskip}
\hline
\noalign{\smallskip}
13435 & 2014-10-16 & $F336W$ & WFC3/UVIS & 1061 & 0$^h$26$^m$49.00$^s$ &$-71\degr32\arcmin10\farcs00$ & $-$115.233
\\
13435 & 2014-10-16 & $F336W$ & WFC3/UVIS & 1061 & 0$^h$26$^m$49.00$^s$ &$-71\degr32\arcmin10\farcs00$ & $-$115.233 
\\
13435 & 2014-10-16 & $F438W$ & WFC3/UVIS & 200 & 0$^h$26$^m$49.00$^s$ &$-71\degr32\arcmin10\farcs00$ & $-$115.231
\\
13435 & 2014-10-16 & $F438W$ & WFC3/UVIS & 200 & 0$^h$26$^m$49.00$^s$ &$-71\degr32\arcmin10\farcs00$ & $-$115.234
\\
13435 & 2014-10-16 & $F814W$ & WFC3/UVIS & 100 & 0$^h$26$^m$49.00$^s$ &$-71\degr32\arcmin10\farcs00$ & $-$115.233
\\
13435 & 2014-05-16 & $F336W$ & WFC3/UVIS & 1061 & 0$^h$26$^m$49.00$^s$ &$-71\degr32\arcmin10\farcs00$ & 84.8355	
\\
13435 & 2014-05-16 & $F336W$ & WFC3/UVIS & 1061 & 0$^h$26$^m$49.00$^s$ &$-71\degr32\arcmin10\farcs00$ & 84.8365
\\
13435 & 2014-05-16 & $F438W$ & WFC3/UVIS & 200 & 0$^h$26$^m$49.00$^s$ &$-71\degr32\arcmin10\farcs00$ & 84.8366
\\
13435 & 2014-05-16 & $F438W$ & WFC3/UVIS & 200 & 0$^h$26$^m$49.00$^s$ &$-71\degr32\arcmin10\farcs00$ & 84.8342	
\\
13435 & 2014-05-16 & $F814W$ & WFC3/UVIS & 100 & 0$^h$26$^m$49.00$^s$ &$-71\degr32\arcmin10\farcs00$ & 84.8342	
\\
14069 & 2016-05-01 & $F343N$ & WFC3/UVIS & 1650 & 0$^h$26$^m$49.00$^s$ & $-71\degr32\arcmin7\farcs99$ & 79.8346
\\
14069 & 2016-05-01 & $F343N$ & WFC3/UVIS & 800 & 0$^h$26$^m$49.00$^s$ & $-71\degr32\arcmin7\farcs99$ & 79.8366
\\
14069 & 2016-05-01 & $F343N$ & WFC3/UVIS & 500 & 0$^h$26$^m$49.00$^s$ & $-71\degr32\arcmin7\farcs99$ & 79.8346
\\
10396 & 2006-01-21 & $F555W$ & ACS/WFC & 496 & 0$^h$26$^m$48.60$^s$ & $-71\degr32\arcmin7\farcs68$ & 289.2675
\\
10396 & 2006-01-21& $F555W$ & ACS/WFC & 496 & 0$^h$26$^m$48.60$^s$ & $-71\degr32\arcmin7\farcs68$ & 289.2700
\\
10396 & 2006-01-21 & $F555W$ & ACS/WFC & 496 & 0$^h$26$^m$48.60$^s$ & $-71\degr32\arcmin7\farcs68$ & 289.2675
\\
10396 & 2006-01-21 & $F555W$ & ACS/WFC & 496 & 0$^h$26$^m$48.60$^s$ & $-71\degr32\arcmin7\farcs68$ & 289.2691
\\
10396 & 2006-01-21 & $F555W$ & ACS/WFC & 20 & 0$^h$26$^m$48.60$^s$ & $-71\degr32\arcmin7\farcs68$ & 289.2704
\\
10396 & 2006-01-21 & $F555W$ & ACS/WFC & 20 & 0$^h$26$^m$48.60$^s$ & $-71\degr32\arcmin7\farcs68$ & 289.2689
\\
10396 & 2006-01-21 & $F814W$ & ACS/WFC & 474 & 0$^h$26$^m$48.60$^s$ & $-71\degr32\arcmin7\farcs68$ & 289.2675
\\
10396 & 2006-01-21 & $F814W$ & ACS/WFC & 474 & 0$^h$26$^m$48.60$^s$ & $-71\degr32\arcmin7\farcs68$ & 289.2691
\\
10396 & 2006-01-21 & $F814W$ & ACS/WFC & 474 & 0$^h$26$^m$48.60$^s$ & $-71\degr32\arcmin7\farcs68$ & 289.2704
\\
10396 & 2006-01-21 & $F814W$ & ACS/WFC & 474 & 0$^h$26$^m$48.60$^s$ & $-71\degr32\arcmin7\farcs68$ & 289.2689
\\
10396 & 2006-01-21 & $F814W$ & ACS/WFC & 10 & 0$^h$26$^m$48.60$^s$ & $-71\degr32\arcmin7\farcs68$ & 289.2675
\\
10396 & 2006-01-21 & $F814W$ & ACS/WFC & 10 & 0$^h$26$^m$48.60$^s$ & $-71\degr32\arcmin7\farcs68$ & 289.2700
\\
\noalign{\smallskip}
\hline
\end{tabular}
\end{table*}

\subsection{Photometry}
\label{subsec:phot}

The ACS/WFC observations were processed through the standard HST pipeline that calibrates for bias, dark and low-frequency flats. Also pixel-based imperfect charge transfer efficiency (CTE) corrections have been applied to the data \citep{AndersonBedin10}. The stellar photometry for NGC~121 was 
derived using the method of point spread function (PSF) fitting, using the spatial variable "effective PSF" (ePSF) libraries for ACS/WFC developed by Anderson \citep{AndersonKing06}. The derived spatial positions of the stars have been corrected for the ACS/WFC geometric distortion 
for each exposure and filter and were matched to 
the longest exposure in $F555W$ employing a linear 
transformation between each coordinate system.
Finally, the instrumental magnitudes from the ePSF measurements were transformed 
into the VEGAMAG system applying the photometric corrections (aperture corrections and zero-points) as described in
\citet{Sirianni05}.

The WFC3/UVIS observations were processed through the standard HST pipeline, as well. The images were also corrected for the imperfect CTE and 
simultaneously calibrated for bias, dark, low-frequency flats and new improved 
UVIS zero-points fully described in \citet{Ryan16}.
Stellar photometry for NGC~121 was derived with PSF fitting methods, using the spatially variable ePSF libraries for each of the WFC3/UVIS calibrated filters developed by Anderson (private communications) which are similar to the ones for ACS/WFC \citep{AndersonKing06}. The instrumental magnitudes from the ePSF measurement were then transformed into the
VEGAMAG system applying aperture corrections 
using bright and well isolated stars on each exposure of the drizzled
images and using the newly derived improved UVIS VEGAMAG zero-points from the WFC3 instrument website\footnote{\url{http://www.stsci.edu/hst/wfc3/analysis/uvis_zpts}}.
Finally, the derived stellar positions were corrected  for the WFC3/UVIS geometric distortion \citep{Bellini11} for each exposure and the positions for each filter were matched to 
the ones from the long exposures in the $F336W$ filter with linear 
transformations between each coordinate system with  
a tolerance of $\sim$0.1 pixels. 

In a final step we combined the photometric sets from ACS/WFC and WFC3/UVIS. For this, the ACS/WFC data set was scaled, 
rotated and then linearly transformed into the coordinates system of the WFC3/UVIS
photometric set using well measured
stars with a matching tolerance of  $\sim0.$1 pixels between two  coordinate systems. The final photometry for each star from each exposure was determined from an average of all measurements in each filter in ACS/WFC as well as in 
WFC3/UVIS, weighted by the quality of the PSF fit. The photometric errors were calculated as RMS deviation of the
independent measurements in the different exposures.

\section{Analysis}
\label{sec:analysis}

\subsection{Structural Profiles}
\label{sec:struc}

\begin{figure}
  \includegraphics[width=\columnwidth]{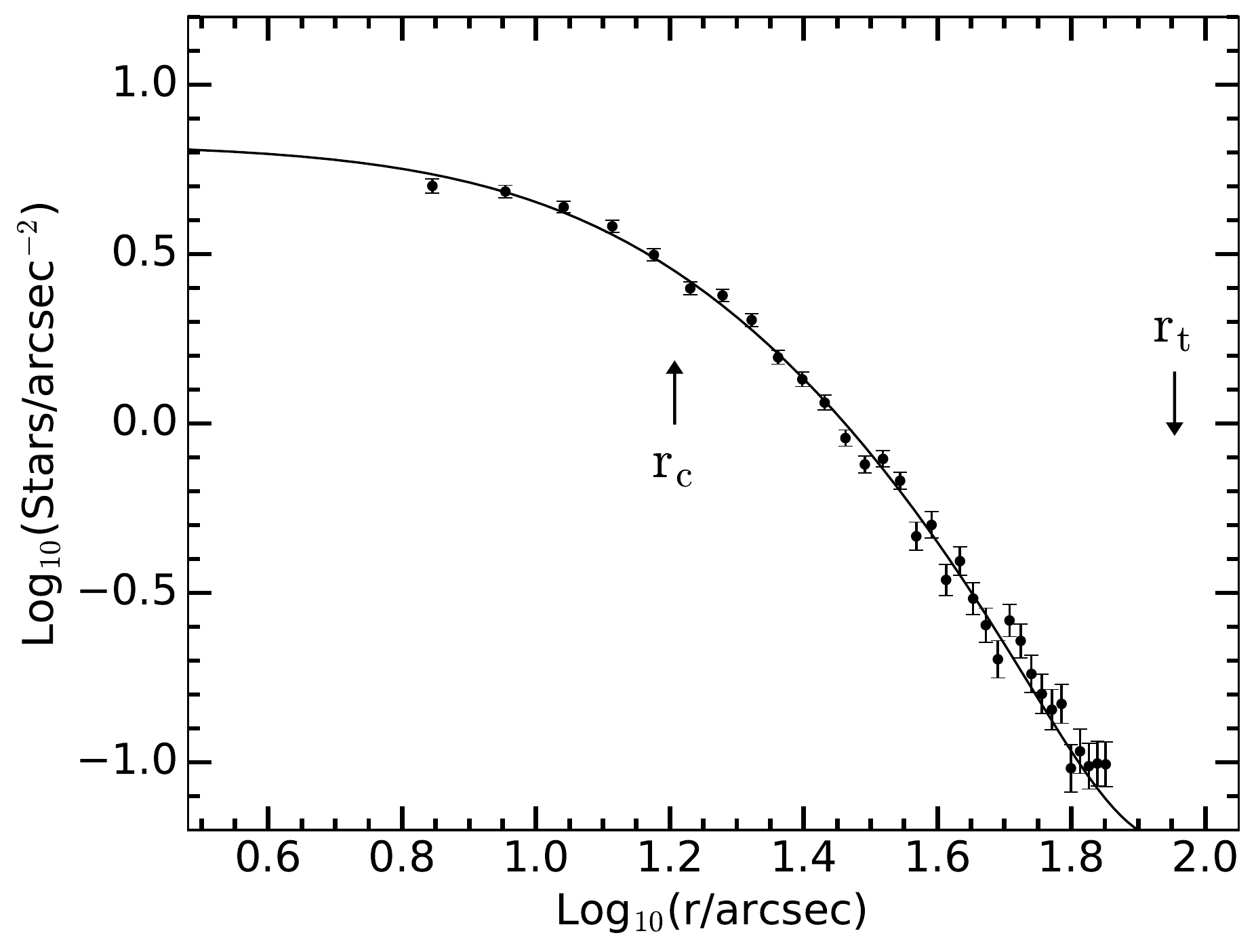} 
  \caption{Radial surface density profile of NGC~121 (individual points with errorbars). We used a binning of 50 pixels (2$\arcsec$) to create the density profile. For illustration, a King profile that was derived from a fit to the discrete stellar positions (see text for more detail) is also shown as a black solid line. The arrows indicate the location of the core radius $r_c$ and the tidal radius $r_t$.}
   \label{fig:density}
\end{figure}

For our analysis we used all stars that are within 2,000 pixels (80$\arcsec$, given the pixel scale of 0$\farcs$04 of WFC3/UVIS) from the centre of NGC~121. 
In order to derive the cluster's parameters we used the discrete maximum likelihood approach outlined in detail in \citet{Martin08} and \citet{Kacharov14}.
We assumed spherical symmetry and fitted an analytic King profile \citep{King62} using the chip coordinates of all stars in the field brighter than 24\,mag in the $F438W$ filter, simultaneously optimizing for 5 free parameters - the centroid of the cluster ($x_0, y_0$), its core and tidal radii ($r_c, r_t$), and a uniform field contamination ($n_f$).
The parameters were iterated in a Markov Chain Monte Carlo (MCMC) manner using the Metropolis-Hastings algorithm \citep{Hastings70}.
We did $10 000$ iterations and adopted the mean and the standard deviations of the last $20\%$ of the Markov chain to be the best-fit values of the free parameters and their uncertainties, respectively. The chain burn-in phase required typically 1500 iterations.
We find a core radius $r_c = 16\farcs1 \pm 0\farcs4$ and and a tidal radius $r_t =  90\farcs0 \pm 3\farcs0$ (see Figure~\ref{fig:density}). At the distance to the SMC, these values correspond to a core radius of 4.9~pc and a tidal radius of 27.3~pc. Our $r_c$ estimate is in good agreement with the value of $r_c = 15\farcs26$ found by \citet{Glatt09} by fitting a King profile to the number density profiles. Our estimate for the tidal radius is however significantly smaller than the value published in \citet{Glatt09}, $r_t = 165\arcsec$.
This discrepancy is not surprising given that the size of the cluster exceeds the observed field. We find a field contamination of  $0.058 \pm 0.004$~stars per square arcsec or roughly $9\%$ of the stars falling in the selection criteria.

\subsection{The Overall CMD}

\begin{figure}
\centering
\includegraphics[width=\columnwidth]{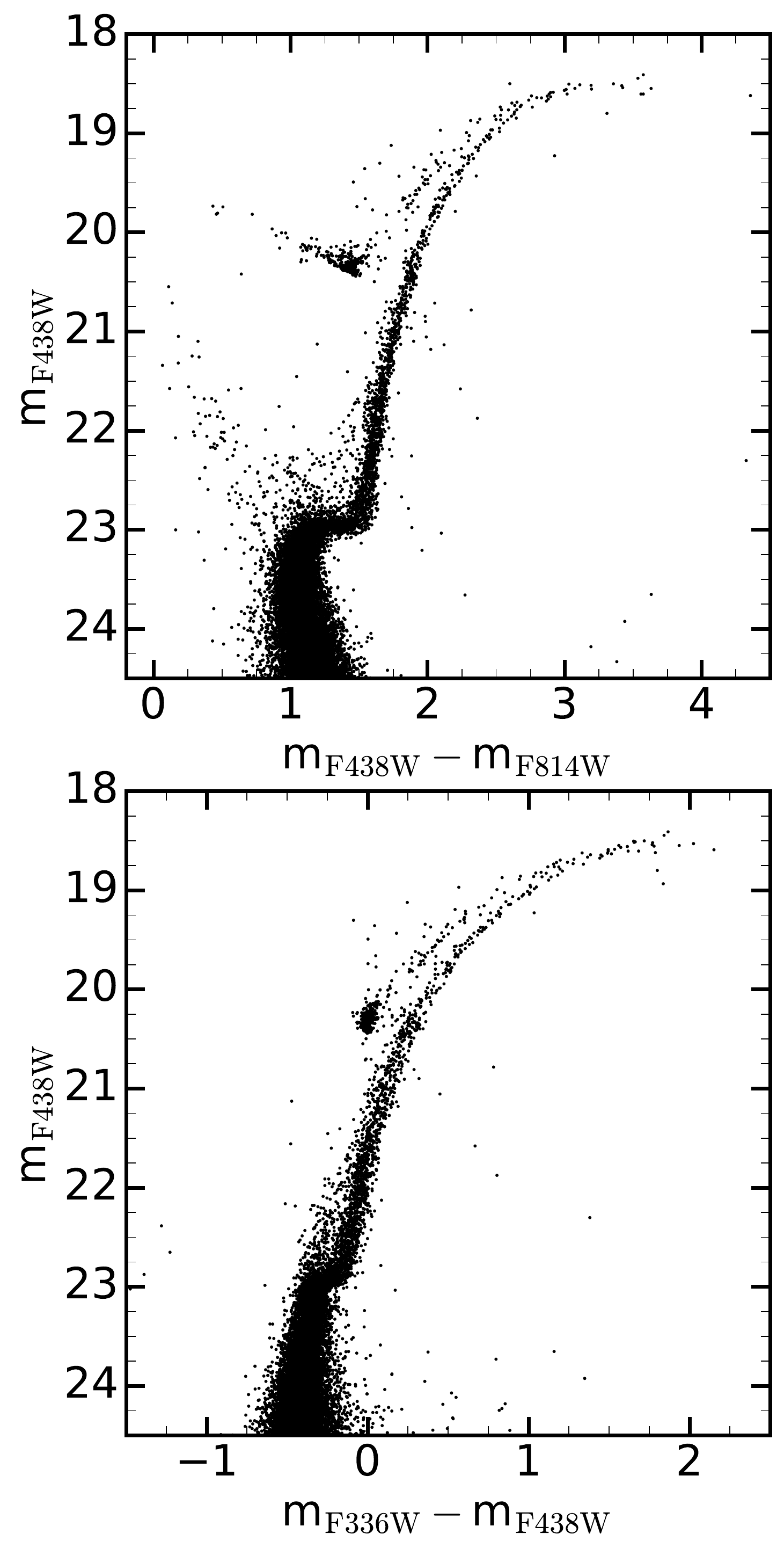}
\caption{Colour-magnitude diagram of NGC~121 using $m_{F438W}$ vs. $m_{F336W}-m_{F438W}$ (bottom panel) and $m_{F438W}$ vs. $m_{F438W}-m_{F814W}$ (top panel) . Shown are all stars within 2000 pixels (80$\arcsec$) from the cluster centre. In the CMD using the $F336W$, the RGB already appears wider than expected from observational errors due to the variations of N in the NH band within the $F336W$ filter (see \citealt{Dalessandro16}).}
\label{fig:ngc121_cmds}
\end{figure}

Figure \ref{fig:ngc121_cmds} shows CMDs of NGC~121 in the $m_{F438W}$ vs. $m_{F336W}-m_{F438W}$ and $m_{F438W}$ vs. $m_{F438W}-m_{F814W}$ space. \citet{Glatt08a} reported an age of 10.5~Gyr for this cluster and a distance modulus $(m-M)$ of 19.06~mag, which is close to the average value of the SMC \citep[18.96~mag,][]{Scowcroft16}.
In the CMD using the $F336W$, the RGB appears wider than expected from observational errors due to the variations of N in the NH band within the $F336W$ filter, which was already shown in the study by \citealt{Dalessandro16}. Using spectroscopic measurements of five RGB stars the authors also found an average iron abundance of [Fe/H] = $-1.28$~dex in NGC~121.

\subsection{The Red Giant Branch}
\label{subsec:RGB}

\begin{figure}
\centering
\includegraphics[width=\columnwidth]{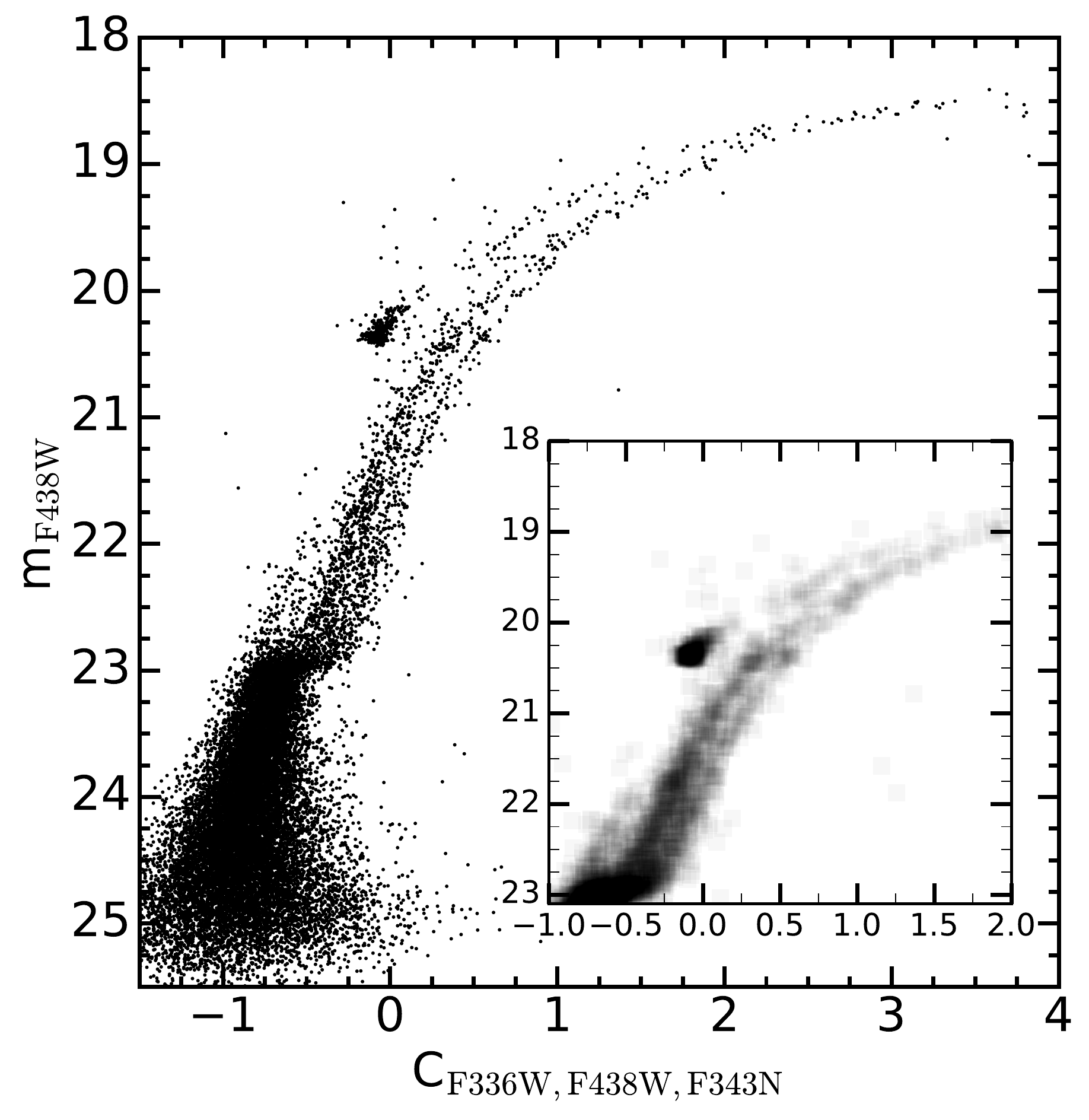}
\caption{$m_{F438W}$ vs. $C_{F336W,F438W,F343N}$ CMD of NGC~121. Using this filter combination, the RGB splits into two branches that are clearly distinguishable. The inlay shows a Hess diagram zooming into the RGB region of NGC~121 illustrating again the split RGB and also revealing the presence of two distinct RGB bumps.}
\label{fig:ngc121_rgb_hess}
\end{figure}

\begin{figure*}
 \begin{tabular}{c c}
  \includegraphics[width=8.0cm]{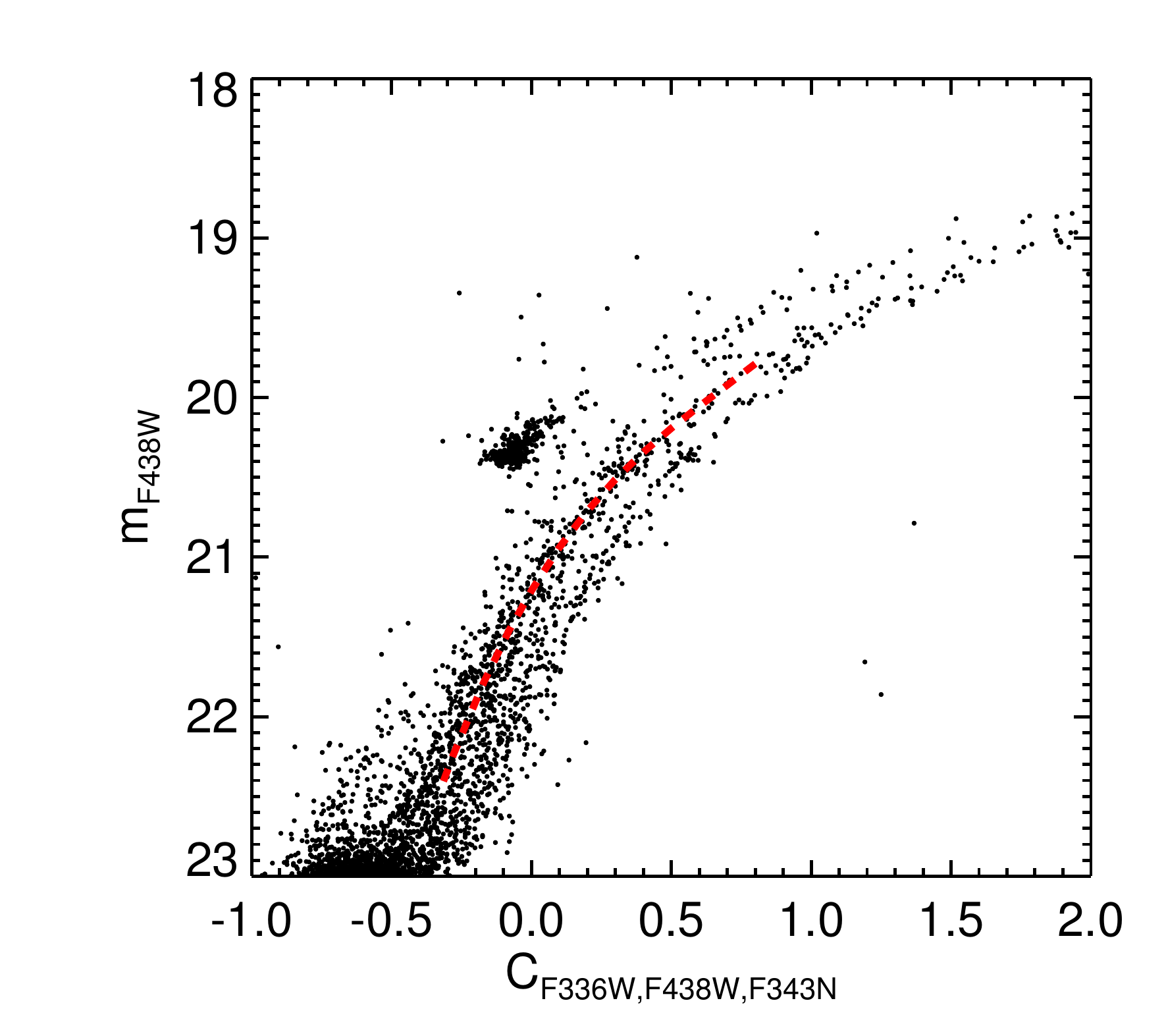} &
  \includegraphics[width=8.0cm]{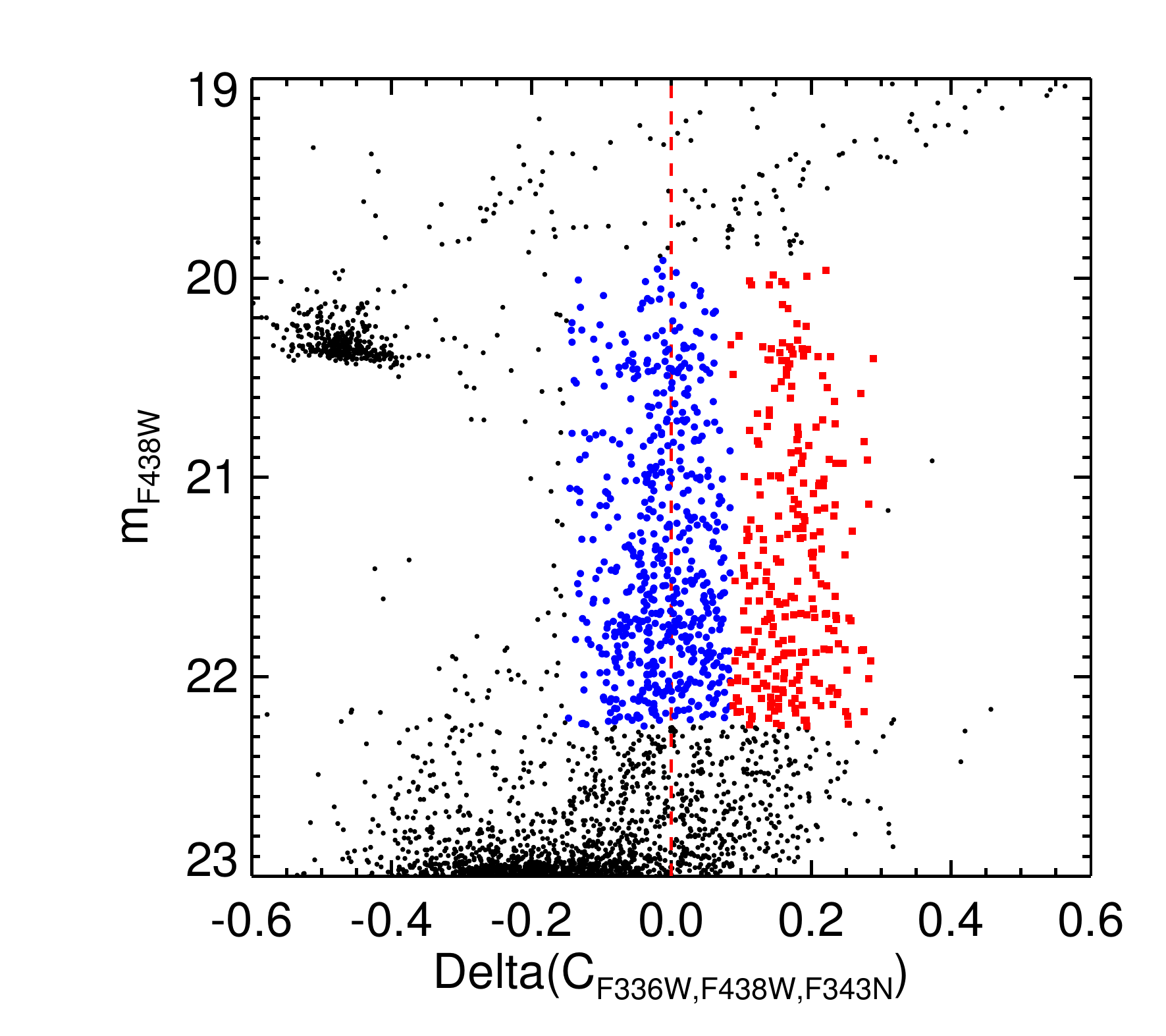} \\
 \end{tabular}
  \caption{\textbf{Left Panel:} Zoom into the RGB region. The dashed red line is the fiducial line to the primordial population. \textbf{Right Panel:} Verticalized RGB, where the x-axis gives the distance to the previously defined fiducial line. The primordial population is colour-coded by blue dots, whereas the enriched population is denoted by red dots.}
   \label{fig:rgb}
\end{figure*}

We now explore the photometry of NGC~121 in the $m_{F438W}$ vs. $C_{F336W,F438W,F343N}$ CMD. We use for the analysis all stars within 2000 pixels (80$\arcsec$) from the cluster centre that have detections in all of the three filters $F336W$, $F438W$ and $F343N$. We did not subtract any contamination of field stars from the CMD. We determined a value of 9\% of field stars across the cluster field (see Section \ref{sec:struc}) and it should be even less at the post-MS part of NGC~121. Therefore, the contribution of unrelated field stars to the RGB part of NGC~121 will not affect our overall results. The resulting CMD is shown in Figure~\ref{fig:ngc121_rgb_hess}. We see that the RGB clearly splits into two discrete sequences, as expected from the models in the case two populations are present in the cluster. The split is most evident in the magnitude range $22.00 \la m_{F438W} \la 19.75$. At brighter magnitudes the two sequences seem to merge again. The brighter/bluer sequence corresponds to stars belonging to the primordial population whereas the fainter/redder sequence consists of the enriched population.

\begin{figure}
  \includegraphics[width=\columnwidth]{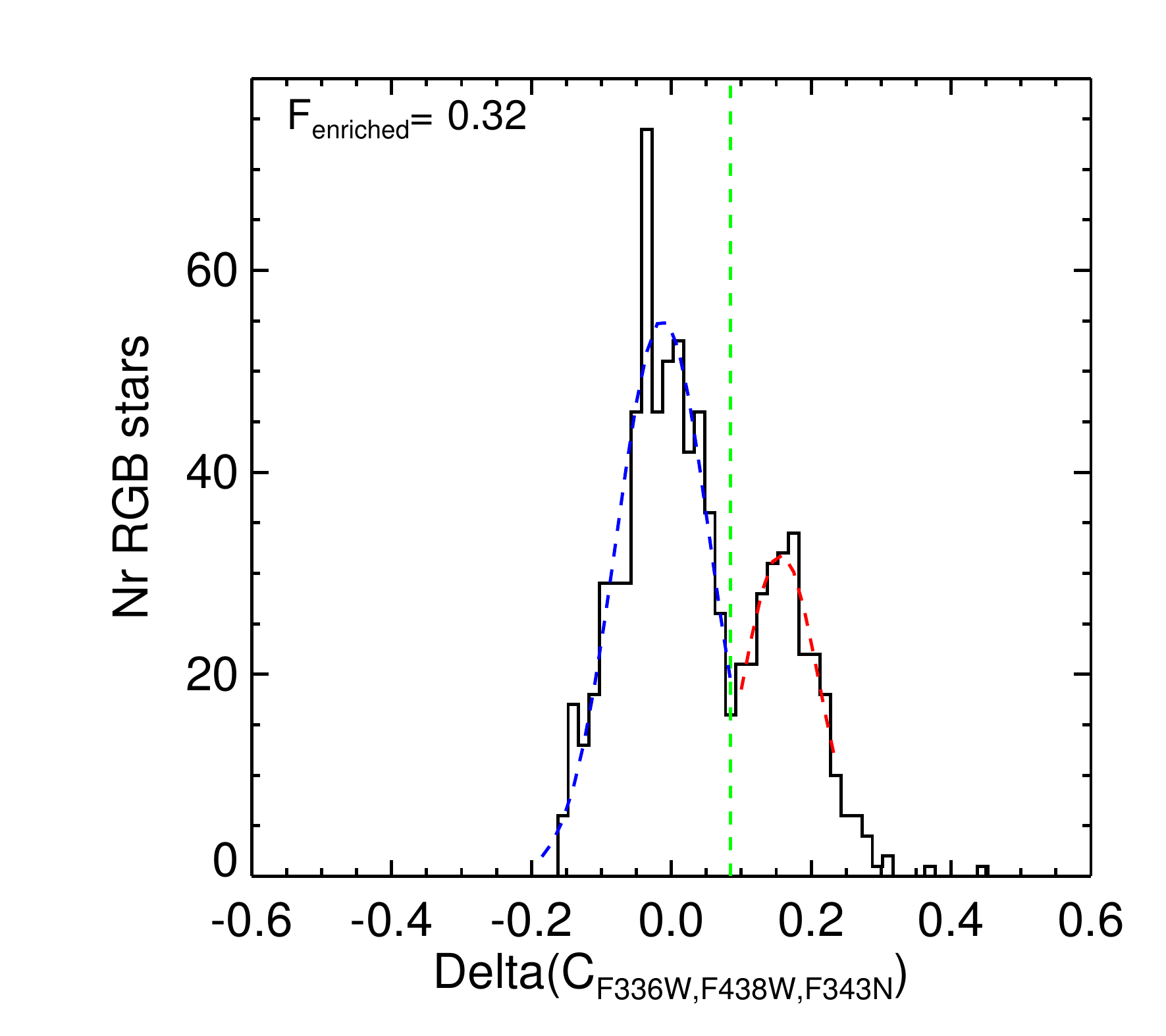} 
  \caption{Histogram of the distribution of the primordial and enriched stars marked in the right-hand panel of Figure~\ref{fig:rgb}. The vertical green dashed line marks the minimum of the distribution which we chose to separate between the two populations. For visualization, the best-fit two component Gaussian function to the distribution is also shown as the dashed line in blue and red. We find a fraction of enriched stars of 32\%$\pm$3\%.}
   \label{fig:hist}
\end{figure}

In the following we analyze the properties of the two sequences in more detail. We start with the determination of the fraction of second population stars. For this, we first defined a fiducial line along the sequence of the primordial stars. This line is a polynomial fit through the points of highest stellar density along the sequence. A zoom into the RGB region of NGC~121 along with the resulting fiducial line as a red dashed curve is shown in the left-hand panel of Figure~\ref{fig:rgb}. We then verticalized the CMD in such a way that the x-axis gives the distance of each star from the fiducial line (Delta($C_{F336W,F438W,F343N}$)). In the right-hand panel of Figure~\ref{fig:rgb} we show the verticalized CMD. The fiducial line is indicated as a vertical red dashed line at Delta($C_{F336W,F438W,F343N})=0.0$. The two populations are evident in the verticalized diagram. To determine the fraction of enriched stars we selected all stars with $ 19.9 \le m_{F438W} \le 22.25$ and $\mathrm{Delta}(C_{F336W,F438W,F343N}) \ge -0.15$. 
Figure~\ref{fig:hist} shows the distribution of the selected stars in form of a histogram (black solid line).
The minimum of the distribution is marked with a vertical green dashed line that lies at Delta($C_{F336W,F438W,F343N})=0.085$. We used this minimum to separate the primordial from the enriched population. By summing up the numbers of stars within the two populations, we found as a result that NGC~121 has a fraction of enriched stars of 32\%$\pm$3\%. This faction is derived from RGB stars within 80$\arcsec$ from the centre, while the relative fraction of pristine vs. enriched stars possibly depends on the actual distance from the centre.

We also tried to estimate the amount of enrichment of the second population from the splitting of the two sequences in the RGB. For this, we fitted a Gaussian to the red sequence in the verticalized CMD using the interval $20.5 \le m_{F438W} \le 21.5$ where the two tracks are nearly parallel. We found the distance to the primordial stars to be $\mathrm{Delta}(C_{F336W,F438W,F343N})$ = 0.17. We then verticalized the isochrones shown in Figure \ref{fig:isoc} and calculated the distance of the intermediate and enriched isochrone to the primordial one in the corresponding magnitude interval. By assuming that the splitting is proportional to the amount of enrichment, we found the following level of enrichment of the second population in NGC~121: the over-abundance in N and Na is about 1.1~dex and 0.5~dex with respect to the primordial population and the level of depletion in O and C is around 0.5~dex and 0.4~dex with respect to the primordial population. Of course, these numbers are only a rough estimate and should indicate a first idea of the order of enrichment in this cluster.

\begin{figure}
  \includegraphics[width=\columnwidth]{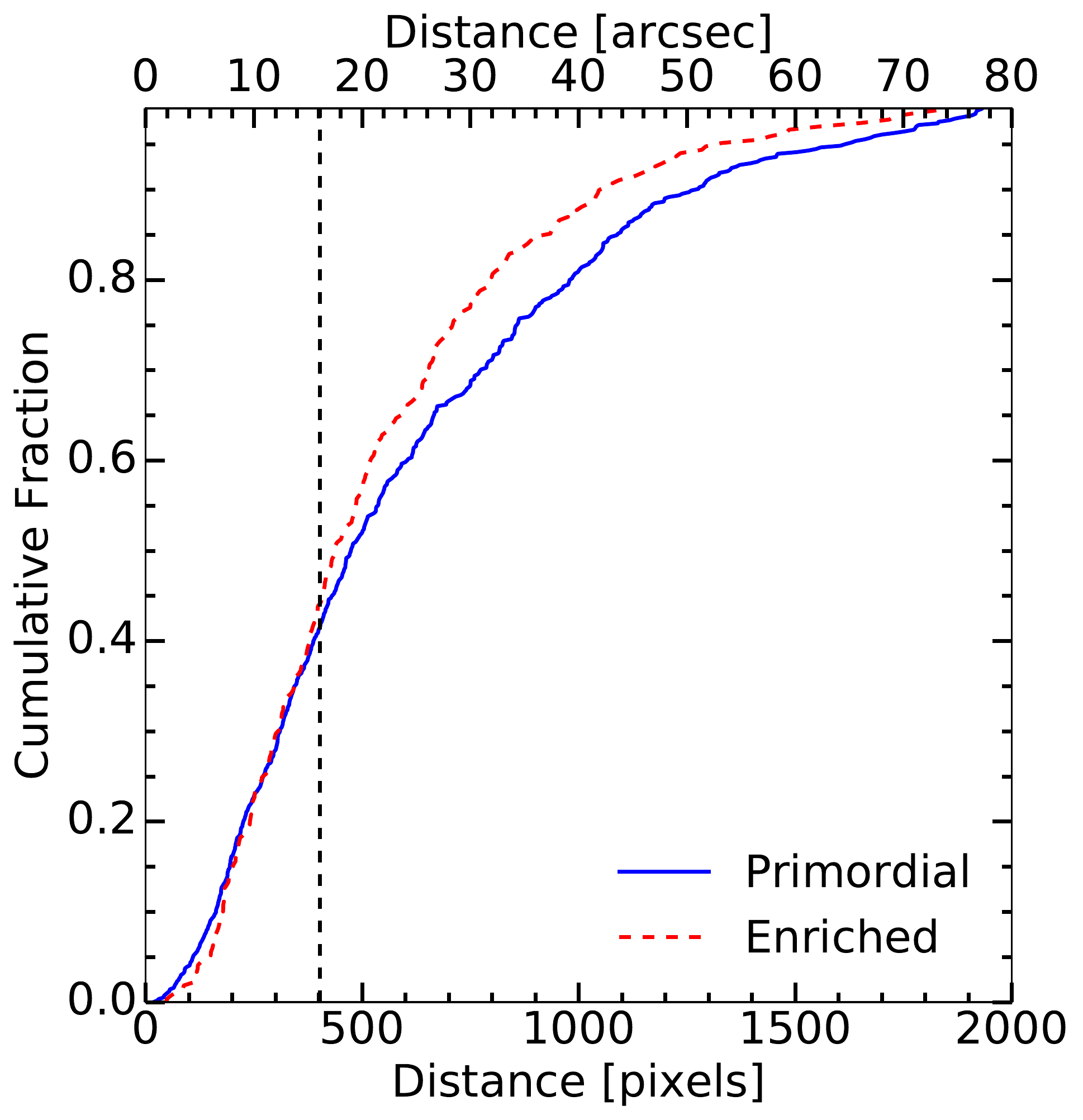} 
  \caption{Cumulative distribution of the two stellar populations as a function of the distance from the cluster centre. The enriched stars (red dashed line) seem to be more centrally concentrated than the stars with a primordial chemical composition (blue solid line). The vertical black dashed line indicates the position of the core radius of NGC~121 at 16$\farcs$1 (403~pixels).}
   \label{fig:cumul}
\end{figure}

Finally, we explore how the two populations are distributed as a function of the radial distance from the cluster centre. We created a radial cumulative fraction distribution of the pristine and the enriched (stars coloured in blue and red in the right-hand panel in Figure \ref{fig:rgb}, respectively) population. For this, we used stars that are within 2,000 pixels from the cluster centre and have pixel y-coordinates smaller than the y-coordinate of the centre of the cluster as only in this region the cluster is complete up to a radius of 2,000 pixels. We found that the two radial distributions seem to agree within the core radius. At larger radii the stars from the second population are more centrally concentrated than first population stars (Figure \ref{fig:cumul}). A KS test gives a probability of 5.0\% that the two samples are drawn from the same underlying distribution. 

Our results for the number ratio of the two populations as well as their radial distribution are in excellent agreement with what was obtained by \citet{Dalessandro16} who used a partially different data set.

\subsection{The Horizontal Branch}
\label{subsec:HB}

\begin{figure}
  \includegraphics[width=\columnwidth]{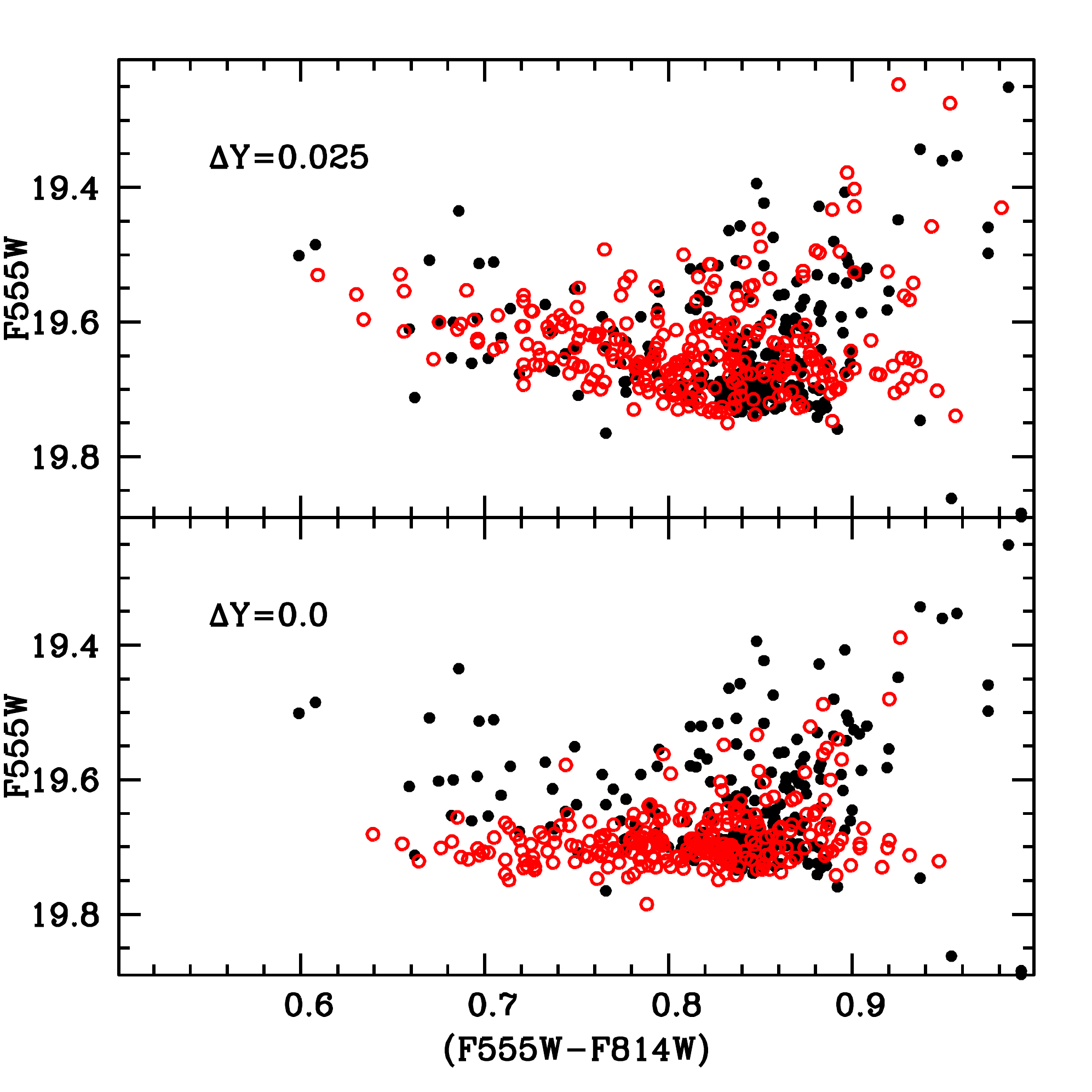} 
  \caption{Comparison of the HB morphology of NGC~121 with synthetic HB simulations. The black dots are the observed data of NGC~121 whereas the red open circles are the models. The top panel shows simulations with a He spread of $\Delta Y$=0.025, ranging from $Y$=0.248-0.273. The bottom panel shows models with a constant He value of $Y$=0.248.}
   \label{fig:HB}
\end{figure}

The presence of multiple populations within this cluster can be inferred also 
from the analysis of its horizontal branch (HB) morphology.
It is well established that together with C-N, Na-O and (sometimes) Mg-Al 
anti-correlations, the various populations hosted by individual globular clusters 
display a range of initial helium abundances \citep[see, e.g.,][for a compilation of results, 
and references therein]{Bastian15b}. In optical CMDs (i.e., 
in photometric bands unaffected by the light-element  anti-correlations) 
of clusters with a red HB morphology, 
a range of initial helium mass fractions $Y$ produces a wedge-shaped 
HB, that cannot be matched by synthetic models 
with constant $Y$ 
\citep[see, e.g., the case of 47~Tuc discussed in][]{dicri, scp16}.
This is precisely also the case in NGC~121, as shown in Figure~\ref{fig:HB},  
that displays a comparison between the observed and an synthetic HB in the $m_{F555W}$ vs $m_{F555W}-m_{F814W}$ CMD,  
calculated with and without a $Y$ spread (the displayed synthetic HBs 
are populated by the same number of objects as the observed HB).

More in detail, we have calculated 
synthetic HB models by employing HB tracks from the BaSTI 
$\alpha$-enhanced stellar model library (\citealt{basti}, and the code  
fully described in \citealt{Dalessandro11,dale}). We made use of  
tracks for [Fe/H]=$-$1.31, [$\alpha$/Fe]=0.4 (corresponding to Z = 0.002), 
and varying $Y$. 
As discussed in \citet{scp16},  
our calculations require to input four parameters, in addition to the cluster initial 
composition, age (we take 10.5~Gyr), 
and photometric error (1$\sigma$ Gaussian error taken from the 
mean photometric error derived for the HB stars). 
Two of these parameters determine the initial $Y$ distribution;  
we considered a uniform distribution with minimum value $Y_{min}$=0.248 
and a range $\Delta Y$.
The other two parameters are the mean value of the mass lost along the RGB, $\Delta M_{RGB}$, 
and the spread around this mean value. 
The adopted cluster age determines the initial value of the mass of the stars 
evolving at the tip of the RGB (denoted as RGB progenitor mass) and translates 
$\Delta M_{RGB}$ value into actual HB masses.

The two synthetic CMDs displayed in Figure~\ref{fig:HB} have been calculated 
with $\Delta Y$=0 and 0.025, respectively. The constant $Y$ HB 
has $\Delta M_{RGB}$ ranging between 0.19 and 0.225~$M_{\sun}$, with a uniform 
probability distribution. A mass loss range is 
in this case necessary to reproduce the colour extension of the observed HB.
The simulation with $\Delta Y$=0.025 has employed $\Delta M_{RGB}$=0.19 $M_{\sun}$ 
for each $Y$, with a very small Gaussian 1$\sigma$ spread equal to 0.005~$M_{\sun}$.
The rationale behind keeping $\Delta M_{RGB}$ constant 
and with a small spread at all $Y$ is that the colour extension of the HB 
is driven mainly by the variation of $Y$ rather than mass loss efficiency 
\citep[see, e.g.,][for more details]{dantona02, Dalessandro11,dale, scp16} because 
populations with the same age and increasing initial $Y$ have a lower 
initial RGB progenitor mass, hence produce increasingly bluer HB objects 
when $\Delta M_{RGB}$ is kept constant\footnote{The adopted age is not 
crucial in our comparisons. A variation by $\pm$ 1~Gyr 
around the reference value causes a change of the 
RGB progenitor mass by about $\pm$0.02~$M_{\sun}$ (lower mass for increasing age). 
Synthetic HBs that match the CMD location of the observed one  
would then require the same HB mass distribution, 
but $\Delta M_{RGB}$ would vary by about $\pm$0.02~$M_{\sun}$ 
(decreased when the age increases) because of the change of the RGB 
progenitor mass.}  

Finally, in the comparison of Figure~\ref{fig:HB} we have employed a reddening 
$E(B-V)$=0.03 as for the comparison with isochrones, and a distance modulus 
$(m-M)$=19.00 (in good agreement with the values found by \citealt{Glatt08a}) obtained by matching 
-- making use of histograms of star counts as a function of $F555W$ magnitudes -- 
the lower envelope of the observed HB with the synthetic ones.

It is clear that the HB calculated with a range of $Y$ matches the observed one better 
than the constant-$Y$ HB. This latter 
has a horizontal lower envelope in the CMD, whereas the observed HB 
lower envelope gets brighter with decreasing colour, and is matched only when a range of 
$Y$ is included in the simulations. The reason is very simple: As mentioned before,  
increasing $Y$ at fixed age means lower RGB progenitor masses and lower --hence bluer-- 
HB masses for a constant $\Delta M_{RGB}$. Given that an increase of initial $Y$ also makes 
models at the start of the HB phase brighter, 
these two effects combined explain naturally why moving towards the blue side of 
the synthetic HB, the lower envelope of the stellar distribution gets increasingly brighter. 

We did not try to enforce a perfect statistical agreement between 
the theoretical and observed star counts, because this rests on the knowledge of 
the accurate statistical distribution of $\Delta M_{RGB}$ and $Y$ among the cluster stars, 
that is at present not available. 
Owing to the lack of theoretical and/or empirical guidance, this distribution may be 
extremely complicated and/or discontinuous. 
The qualitative constraint we have imposed on the matching synthetic HB is  
however sufficient to establish the presence of a range $\Delta Y$, hence of multiple 
populations also on the HB.
Increasing $\Delta Y$ to 0.03 or above, or decreasing this range below 0.02, make the 
synthetic HB too much, or not enough, tilted towards brighter magnitudes when moving to 
bluer colours, respectively. Hence $\Delta Y$ appears to be in the range between 0.02 and 0.03, 
in line with the values obtained for several Galactic globular clusters 
\citep[see][and references therein]{Bastian15b}.

\section{Discussion and Conclusions}
\label{sec:conclusions}

Using a combination of three blue/ultraviolet filters we were able to detect two distinct stellar populations in the RGB of the SMC cluster NGC~121 (however, more populations might be present). The brighter/bluer sequence corresponds to a population with a pristine chemical abundance whereas the fainter/redder sequence is composed of chemically enriched stars. Our findings are in agreement with the recent results by \citet{Dalessandro16}.
They found that the RGB is broader than expected from photometric errors in the $m_{F336W}$ vs $m_{F336W}-m_{F438W}$ CMD. Using a combination of the filters $F336W$, $F438W$ and $F814W$, they detected a splitting in the RGB, as well. We find that the fraction of the second population stars is 32\%, consistent with the results from \citet{Dalessandro16} who found a fraction of enriched stars of 35\%, , using the pseudo-color $C_{F336W,F438W,F814W}$. This, however, is much smaller than the expected median value found in Galactic GCs. \citet{BastianLardo15} collected a sample of 33 GCs and found that the fraction of enriched stars is never smaller than 50\% with median value of 68\%$\pm$7\%. Moreover, this fraction seems to be independent of the cluster mass, metallicity or distance to the centre of the Galaxy. The data set presented in this study is mainly based on spectroscopic data that only probe the outer regions of the clusters. Thus, the result refers to the fraction measured at larger radii. Given our results, the fraction of enriched stars appears to vary from cluster to cluster, much more than reported by \citet{BastianLardo15}. This is consistent with the results of Lardo et al. (in prep.) who found larger variations in Galactic GCs using photometric surveys. Their data are based on much larger statistical samples of stars within individual clusters and also sample the inner regions of GCs. The data sets of \citet{BastianLardo15} and Lardo et al. (in prep.) therefore sample different regions of clusters with different properties and possibly varying ratios of first and second population stars. 

Even though the relative number of enriched stars we find in NGC~121 is low, it is still in tension with scenarios that invoke strong cluster mass loss to go from initial fractions of enriched stars of $\sim$5\% to higher values by preferentially removing stars with primordial chemical composition. If we assume that only first population stars have been removed it follows that NGC~121 must have lost $\sim$90\% of its initial mass in order to get to the observed fraction of 32\% enriched stars. This is, however, only a lower limit as we assume that only stars with a primordial composition have been lost. This high number, however, seems to be unlikely given the weak tidal field of the SMC and the present-day mass of NGC~121 (log(M/M$_{\sun}$)=5.57, \citealt{McLaughlin05}). It is expected that the time it takes to dissolve a star cluster is longer for more massive clusters and in weaker tidal fields \citep[see e.g.][and references therein, for a discussion]{BastianLardo15}. Quite extreme assumptions have to be made for the cluster and its environment in order to allow for such high dissolution rates (see \citealt{D'Ercole08}). Such high cluster dissolution rates are also in tension with observations of the Fornax Dwarf Spheroidal galaxy \citep{Larsen12}.
In contrast, \citet{Kruijssen15} showed in his model for the origin and evolution of GCs that typically, GCs could have only been, on average, at the most a factor of three more massive at birth.

Additionally, we analyzed the radial distribution of the two populations. We found that up to a radius of 300 pixels (12$\arcsec$), which is approximately the core radius of NGC~121, the two populations are distributed the same. Only at larger distances, the second population stars seem to be more centrally concentrated than the primordial stars. In the self enrichment scenario where a second generation of stars forms within a cluster out of a mixture of processed stellar material and pristine gas, it would be expected that this second generation is formed in the centre of the cluster as the gas densities are highest there.  A more centrally concentrated second population would be in agreement with the prediction from this scenario. Other studies from the literature do not provide definitive answers regarding the relative radial distributions of populations in different clusters. Using ground-based photometry or spectroscopy measurements, the enriched populations are generally found to be more concentrated \citep[e.g.][]{Carretta10b, Beccari13, Larsen14, Li14}. But due to the crowding in the inner regions these studies usually avoid the central parts of the clusters. Recently, \citet{Larsen15} analyzed the GC M15 using HST/WFC3 data and found that stars with primordial chemical composition are more centrally concentrated than stars with enhanced N abundances, taking also the central parts of the cluster into account. This trend, however, seems to invert at larger radii \citep{Lardo11}. \citet{Dalessandro11} studied the radial distributions of the two populations found in the SGB and the RGB of NGC~6362, but did not find any significant difference along the two populations across the extent of the cluster. Similarly, \citet{Nardiello15} found that the populations of the red and blue MS in the two GCs NGC~6752 and NGC~6121 (M4) show no difference in their radial distributions.

The results in this paper along with the findings of \citet{Dalessandro16} add the SMC to the list of galaxies (including the Milky Way, e.g. \citealt{Gratton12}, the LMC, \citealt{Mucciarelli09} and the Fornax dwarf spheroidal galaxy, \citealt{Larsen14}) harbouring a GC with multiple populations. Therefore, it appears that this is a ubiquitous property of old GCs independent of environment or galaxy type. However, it is not clear yet what parameter controls whether a star cluster is able to host multiple populations. The scenarios that invoke self-enrichment and multiple episodes of star formation require the clusters to have high masses at birth in order to retain the processed stellar ejecta. 
As NGC~121 is relatively young, compared to Milky Way GCs, formation scenarios that include Pop III stars can already be ruled out. In forthcoming papers we will continue the study of a variety of massive clusters with a range of different ages and masses within the Magellanic Clouds aiming to constrain the parameter that is responsible for the formation of multiple populations in star clusters. 

In the present work, we introduced an ongoing photometric survey using the HST searching for multiple populations in LMC/SMC clusters spanning a large range of ages. We presented, as first results of this survey, the detection of two populations in the RGB of the 10.5~Gyr old SMC cluster NGC~121 as well as evidence of an He spread from the morphology of its HB. In the future, our survey will be capable to provide important observational constraints on the origin of multiple populations by helping to constrain the range of ages and/or masses where they are present.

\section*{Acknowledgements}

We, in particular F.N., N.B. and V.K.-P., gratefully acknowledge financial support for this project provided by NASA through grant HST-GO-14069 from the Space Telescope Science Institute, which is operated by the Association of Universities for Research in Astronomy, Inc., under NASA contract NAS526555.
N.B. gratefully acknowledges financial support from the Royal Society (University Research Fellowship) and the European Research Council (ERC-CoG-646928, Multi-Pop). D.G. gratefully acknowledges support from the Chilean BASAL Centro de Excelencia en Astrof\'{i}sica y Tecnolog\'{i}as Afines (CATA) grant PFB-06/2007. 
M.J.C gratefully acknowledges support from the Sonderforschungsbereich SFB 881 "The Milky Way System" (subproject A8) of the German Research Foundation (DFG).
We are grateful to Jay Anderson for sharing with us his ePSF software.
We thank the anonymous referee for useful comments and suggestions.











\bsp	
\label{lastpage}
\end{document}